\pdfoutput=1
\documentclass[12pt]{iopart}
\usepackage{color, soul, iopams}
\usepackage[pdftex]{graphicx}

\newcommand{\fig}[1]{Fig.~\ref{#1}}
\begin{document}

\title{Limits of Elemental Contrast by Low Energy Electron Point Source Holography}
\author{Lucian Livadaru$^{1,2}$, Josh Mutus$^2$ and Robert A Wolkow$^{1,2}$}
\address{$^1$ National Institute for Nanotechnology, National Research Council of Canada, Edmonton, Alberta T6G 2M9, Canada}
\address{$^2$ Department of Physics, University of Alberta, Edmonton, Alberta T6G 2J1, Canada}
\ead{lucian@ualberta.ca}



\begin{abstract} 
Motivated by the need for less destructive imaging of nanostructures, we pursue point-source in-line holography (also known as point projection microscopy, or PPM) with very low energy electrons ($\sim$100 eV). This technique exploits the recent creation of ultrasharp and robust nanotips, which can field emit electrons from a single atom at their apex, thus creating a path to an extremely coherent source of electrons for holography. Our method has the potential to achieve atom resolved images of nanostructures including biological molecules. We demonstrate a further advantage of PPM emerging from the fact that the very low energy electrons employed experience a large elastic scattering cross section relative to many-kiloVolt electrons. Moreover, the variation of scattering factors as a function of atom type allows for enhanced elemental contrast. Low energy electrons arguably offer the further advantage of causing minimum damage to most materials. Model results for small molecules and adatoms on graphene substrates, where very small damage is expected, indicate that a phase contrast is obtainable between elements with significantly different Z-numbers. For example, for typical setup parameters, atoms such as C and P are discernible, while C and N are not.
\end{abstract}


\maketitle
\date{today}

\section{Introduction}
Currently, numerous fields of research require detailed knowledge on the composition and structure of nanoscale systems. Most techniques operating at the nanoscale target either structure or composition, but due to great challenges, rarely both. A short review of the fundamental and practical difficulties involved is found in Ref.\cite{billinge2}. Recent efforts oriented toward the highly desirable task of simultaneous determining the structure and chemical composition at the nanoscale have seen some success \cite{muller2008asc, findlay2008atomic,klenov2006contributions,pennycook2002zci,kimoto2007esi}. Various forms of electron microscopy (EM) have advanced significantly in recent years  by aberration correction techniques\cite{mcbride2004acz,pennycook2006mat,varela2005mca,muller2009sab,hutchison2005vda,falke2005rsc,tanaka2004foc}, exit-wave reconstruction\cite{allen2004exit,hsieh2004resolution,zandbergen2000exit,oxley2001phase}, and combined techniques\cite{tillmann2004sac}, to mention just a few.

Electron holography has the potential of determining both structure and elemental composition at the nanoscale, partly due to the its capability of detecting changes in the phase of an electron wave by exploiting interference phenomena. In a broader sense, recent progress in exploring the interference properties of electrons are reviewed in Ref.\cite{hasselbach2010progress}. By using interference, holography can also image electric and magnetic fields\cite{cumings2002electron,matteucci1991electron,mccartney2001magnetic,bonevich1993electron}, and carry out important tests of quantum mechanics\cite{tonomura1986evidence}. As such, holography has been pursued under various forms, using both high energy\cite{midgley2001introduction,lichte2008electron,lichte2007electron,de1993detection,cowley1993electron} and low energy electrons\cite{morin1996low,barton1988photoelectron,spence1999introduction}.  

Currently, the imaging of low Z-number materials and biological samples by conventional EM\cite{meyer2008imaging,girit1} or holography\cite{simon2008electron,xu2001digital} is faced with further challenges due to a greater susceptibility to radiation damage, which can become a major contributor to the resolution limit. For the increasingly important case of graphene structures, electron microscopy at high energy presents a peculiar set of challenges, requiring lowest available voltage ($\sim$ 80 keV), non-routine implementation of EM and careful interpretation of the results\cite{meyer2008direct}. Contrast interpretation in the conventional and aberration corrected EM system can also be problematic\cite{wang14}. In this light, alternative techniques employing ultralow voltage ($\sim$ 100 eV) would be helpful for improving the stability of graphene sheets during imaging against radiation-induced defects\cite{girit1}, morphologic changes, etc. Combining the benefits of lower voltage (reduced damage) with those of holography (a more reliable image interpretation) is then an obvious choice for pursuing the above goals.

Here we show that low-energy electron point source (LEEPS) holography, or PPM, has the potential to resolve both the three-dimensional structure\cite{gabor1948new} and the composition of nanoscale systems with atomic resolution. In LEEPS holography, an ultrasharp tip field emits electrons from a single atom \cite{pitters2006tungsten,binh1992electron,silverman1995brightest,binh1997field,cho2004quantitative} towards a weakly scattering nano-sized object. Most of the electron wave passes through the sample unimpeded (the reference wave), while a small amount is scattered by the sample. The reference and scattered waves interfere at a screen to create a pattern that contains a wealth of information about the sample. This pattern can then be inverted to extract -- within some practical limits -- the three dimensional structure of the sample\cite{kreuzer1992theory,livadaru2008reconstruction} and, in some cases, its composition. The latter is usually referred as elemental, or Z-contrast and -- in our method -- becomes available because the subtle information stored in the electron wave phase, after being scattered by a sample, is collected in the interference pattern on the detector. A recent review on a variety of applications of LEEPS microscopy can be found in Ref.\cite{beyer2010low}.

In this work we investigate to what limits the information stored in an electron wave used in LEEPS holography can be practically retrieved. To this aim, we undertake a theoretical study for the simulation in atomistic detail of LEEPS holograms and their reconstruction, given a set of parameters for an experimental setup. We investigate the possibility of obtaining both amplitude and phase reconstruction for nanosized structures with ordered lattice (e.g. graphene), as well as disordered structures (e.g. defects in lattices, adsorbed atoms and molecules, etc.). 

We would like to emphasize the advantages of LEEPS holography for the key application of imaging small molecules (adsorbed on graphene), thus avoiding the damage and morphologic changes inevitable at high energies\cite{meyer2008imaging}. The knock-on damage would be greatly reduced due to much lower electron energy, while damage due inelastic processes is reduced by  efficient electron transport through the graphene substrate. We carefully address the conditions in which atomically resolved elemental contrast is obtainable in this kind of experiment.

\begin{figure}
\centerline{
\includegraphics[width= 0.6\linewidth]{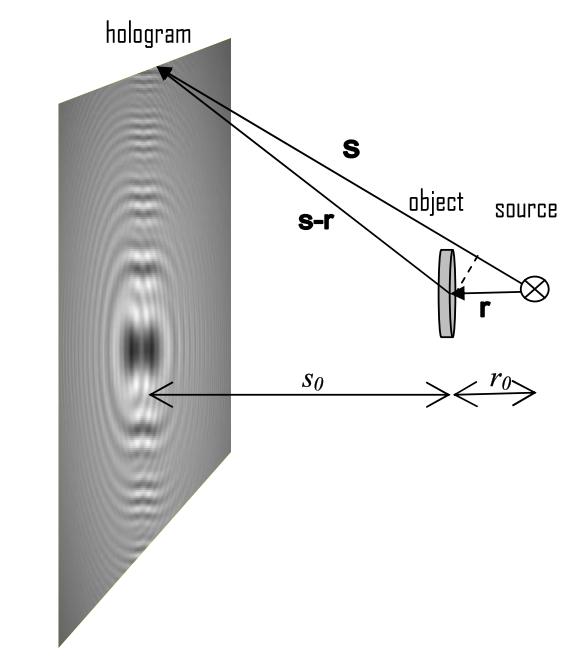}
}\caption{Schematic diagram of the in-line holographic setup showing the source (centre-crossed circle), object, and detector screen (hologram). Position vectors $\bi{r}$ and $\bi{s}$ have the origin at the source and point at an atom in the sample and at a recording pixel on the detector respectively. The vector $(\bi{s} - \bi{r})$ is used in the digital reconstruction algorithm.} 
\label{fig1}
\end{figure}

\section{The in-line holographic method}
An in-line holographic setup is depicted in \fig{fig1}.  An ultrasharp tungsten nanotip is negatively biased with respect to the conducting sample, and it can field emit electrons from a very small area, practically from a single atom at its apex\cite{cho2004quantitative,pitters2006tungsten}.  Field emitted electrons, having a very narrow energy distribution, scatter elastically on the sample and the resulting scattered wave interferes at the detector with the reference electron wave (unperturbed by the sample). The ensuing pattern at the detector is an in-line hologram.

Practically, the in-line holographic experiment requires the existence of a well defined reference wave. In a transmission holographic setup, this is achieved if the object scatters only a small fraction of the incident wave, that is, if the scattering cross section of the object is small compared to the cross section of the electron beam. This condition is in contrast with the diffractive regime, where the scattered wave must be maximized. Also, the success of the holographic experiment (including phase retrieval) depends on the reduction of distortion effects of magnetic fields (chiefly via the magnetic Aharonov-Bohm effect) on hologram formation\cite{livadaru2008line}.

In order to simulate the formation of in-line holograms, we use the scattering theory in its integral form, i.e. Lippmann-Schwinger equation\cite{kreuzer1992theory}. In this theoretical frame, the total wave function at point $\bi{s}$ on a detector placed at an asymptotically large distance from the sample is given by

\begin{eqnarray}\label{eq1}
 \psi_{{\rm tot}} (\bi{s})
	&=& A_{{\rm ref}} (\bi{s})\frac{e^{iks} }{s} +\sum _{j=1,N_{at} }A_{{\rm ref}} (r_{j} )\frac{e^{ikr_{j} } }{r_{j} } \frac{e^{ik|\bi{r_{j}} - \bi{s}|}}{k|\bi{r_j} - \bi{s}|} \nonumber	\\	&&
\times \sum _{l=0,N_{l} }(2l+1)\sin \delta _{l} \exp (i\delta _{l} )P_{l} (\cos \theta _{j} )  
\end{eqnarray} 
where $\bi{s}$ is the position vector on the detector (see \fig{fig1} for vector diagram), $\bi{r_j}$ are the positions of the atomic centers in the scattering object (shown as a generic $\bi{r}$ in that figure), $l$  is the angular momentum number for each partial wave,   $P_l$   are the Legendre polynomials,  $\theta_{j}$   are the scattering angle at each atom, and  $\delta_{l}$  are the scattering phase shifts for each value of $l$. The values of $\delta_{l}$ are assumed to be well approximated by the free atom phase shifts, available for example from the NIST Standard Reference Database 64 \cite{jablonski2003nist}. Empirically, we found that a finite number of partial waves ($N_{l} > 10$) is sufficient to produce the value scattered wave within a good accuracy.

In this study, we simulate holograms produced by a single atom point source. The virtual (incoherent) source size is chosen to be 0.8{\AA} in diameter, which is at the lower bound of previous estimates for emission from tungsten nanotips\cite{cho2004quantitative,chang2009fully}. For any experimental case, incoherent emission of electrons from a nanotip results in an undesirable blurring (overlap of many interference patterns from different wavelengths) of the recorded hologram.  
We do not include in our simulated hologram formation some of the factors that in practice have a degrading effect on the quality of holograms such as inelastic electron scattering, mechanical and electromagnetic noise in the apparatus, secondary electron generation. These are specific to sample, source, and exact experimental apparatus and its environment and as such require all these specifics to be know and then simulated, which is a task outside the scope of this paper. Other researchers have looked at the effects of the coherence of the beam, elastic and inelastic mean free paths and other factors on the resolution of the method, for specific nanotip (source) shapes, specific setup, and a specific sample (DNA). They found that the most severe limiting factor for that setup was a small beam divergence angle of the beam yielding a relatively large effective source size of about 8{\AA} and a resolution limit of 15{\AA}.\cite{stevens2009resolving}
However, as we mentioned above, it was found experimentally that the coherence properties of tungsten nanotips can be improved well beyond these estimates\cite{cho2004quantitative,chang2009fully}, therefore the findings of Ref.\cite{stevens2009resolving} cannot be looked at as fundamental limits of the resolution.
Therefore in our simulations, we sum up the coherence properties of the electron beam in a single parameter namely the virtual source size. Assessing the size of the virtual source size for field emission from a single atom is a challenging task as the exact result depends on the exact shape of the nanotip, as well as the orientation of the terminal atomic layers. Another contribution to the virtual source size comes from the level of electromagnetic noise in the microscope environment. As we aim to emphasize the possibility of elemental contrast in our samples, we assume that our virtual source size is small enough to achieve a high enough spatial resolution, and take the above empirical value of 0.8{\AA}, according to recent experimental findings.

\begin{figure}[tb]
\centerline{
\includegraphics[width= 1.0\linewidth]{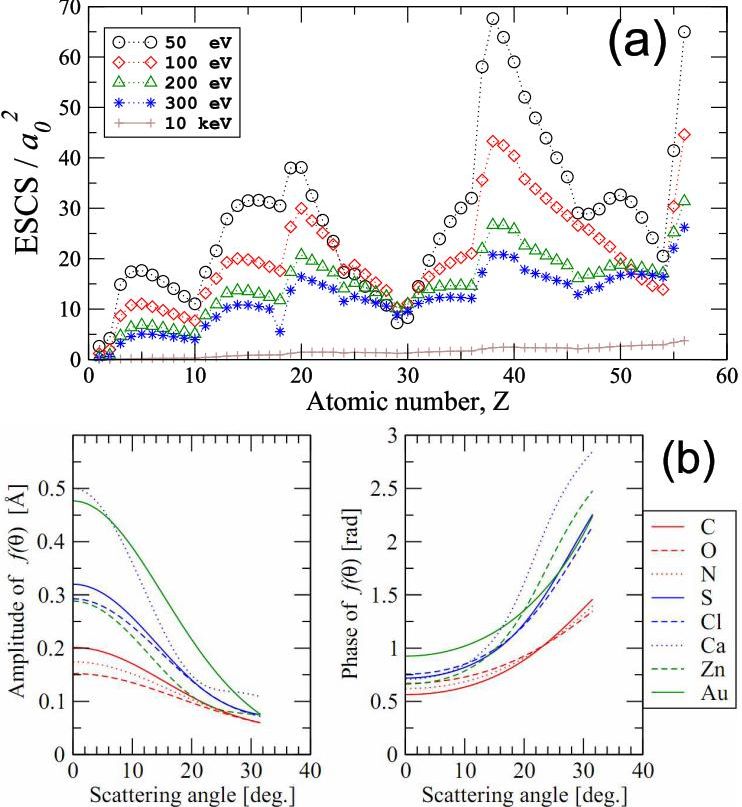}
}
\caption{(a) Total elastic scattering cross sections as a function of the atomic number for different values of electron energy (indicated in the legend). (b) Amplitude and phase of atomic scattering factors as a function of the scattering angle for selected elements indicated in the legend.} 
\label{fig2}
\end{figure}

The reconstruction of in-line holograms can be performed in the frame of Fresnel-Kirchhoff theory of diffraction\cite{livadaru2008line}, according to which the reconstructed wave function is

\begin{equation}
\label{eq2}
\psi _{rec} (\bi{r})=\int_{\rm screen}{\rm d}s_{x} {\rm d}s_{y} I_{\rm holo} (\bi{s})\frac{e^{iks} }{s} \frac{e^{-ik|\bi{r-s}|} }{|\bi{r-s}|}      
\end{equation}
The mechanism for elemental contrast in reconstructed holograms can be explained by realizing the fact that different atomic species exhibit different (complex-valued) scattering factors:

\begin{equation}
f(\theta )=\frac{1}{2ik} \sum _{l}(2l+1)\sin \delta _{l} \exp (i\delta _{l} )P_{l} (\cos \theta )] 
\end{equation}

The total electron elastic-scattering cross section (ESCS) can be obtained by integrating the square of the scattering factor over the solid angle

\begin{equation}
\sigma =2\pi \int _{0}^{\pi }d\theta \sin \theta  |f(\theta )|^{2} =\frac{4\pi }{k^{2} } \sum _{l}(2l+1)\sin ^{2} \delta _{l}   
\end{equation}
Thus, for a given scattering angle, the amplitude and the phase of $f(\theta)$ have values specific to each atomic species. Furthermore, these values can be in principle predicted by quantum mechanical calculations. Thus, by comparison between the reconstructed scattering factors (amplitude and phase) and the ESCS of different atoms in the sample with the theoretically predicted scattering factors, one can obtain spatially resolved chemical information about a sample.  

\section{Results}
In \fig{fig2} (a), we show the ESCS for a range of atomic species, with atomic number up to $Z=56$, and selected values of electron energies. There is a highly non-monotonic variation of this quantity with the atomic number, which in theory opens up the possibility of discriminating the atomic species on this basis. As well, the rate of variation of the ESCS with the electron energy offers another discriminating parameter. Furthermore, as shown in \fig{fig2} (b), the variation of the atomic scattering factor with the scattering angle can also be used as a clue to atomic species in the sample. On average the phase difference for the light and heavy elements considered here becomes greater at higher scattering angles, where the high resolution details are stored, while the opposite holds for the scattering amplitudes.

On careful examination, the variation of the ESCS versus $Z$ can be roughly linked to the electronic configuration of each element. For the first three periods in the periodic table, we notice that local minima in a curve appear at the end of each period, followed by a sharp increase. However, close to the beginning of period 4 starting with Sc ($Z=21$) and the occupation of the $3d$-subshell, this trend is broken and the ESCS experiences a decrease for consecutive elements up to but exclusive of Zn ($Z=30$). This sequence corresponds to the complete filling of the $3d$-subshell with 10 electrons. From Zn on, an upward trend is taken up to the end of the period, corresponding to the filling of the $4p$-subshell. So at the end of this period there is not a minimum as for the preceding periods, but a shoulder followed by a sharp increase and a maximum for Sr ($Z=38$). In period 5 again, with the occupation of $4d$-orbitals a downward trends sets in and the period ends in a local minimum, just like for the first three periods. However, there is more structure inside period 5 as there are now two maxima corresponding to atomic numbers 38 and 50 (\textit{i.e.} to the $5s$ and $5p$ subshells being filled with exactly 2 electrons) and one minimum at $Z=47$ (just before the $4d$-level is filled). Further investigation is needed in order to discriminate to what extend these trends are artifacts of the calculation model used by the NIST database, 
which employs a local central relativistic (Dirac) potentials for all atom types.

At this point, without a careful analysis of the sensitivity and the signal-to-noise ratio of the method, it's difficult to say to what extent the above forms of elemental discrimination is practical within our holographic method. As we said above, beyond the fundamental physical factors accounted in our simulations, other practical limiting factors depend on the sample and on the particulars of the experimental apparatus and its environment. We defer more detailed studies of such particular cases to future endeavors.

Note that the much higher values of the ESCS at low energies ($<$500 eV) as compared to at higher energy ($>$10 keV) shown in \fig{fig2} (a) is the chief reason for which our method is suitable for imaging low-Z materials.

Ideal samples for our microscope are very thin, e.g. graphene-like structures, which allow a strong reference wave (through and around the sample). Atoms and molecules adsorbed on graphene sheets or at the edges of graphene nanoribbons can also be resolved and identified. In order to determine the degree to which elemental contrast is feasible for such systems, we undertook simulation of the hologram formation and subsequent reconstruction of the hologram. This is meant to determine the theoretical limits of the LEEPS holography method. Hologram simulations and reconstructions are done by using the above equations (\ref{eq1}) and (\ref{eq2}).

\begin{figure}[tb]
\centerline{
\includegraphics[width= 1.0\linewidth]{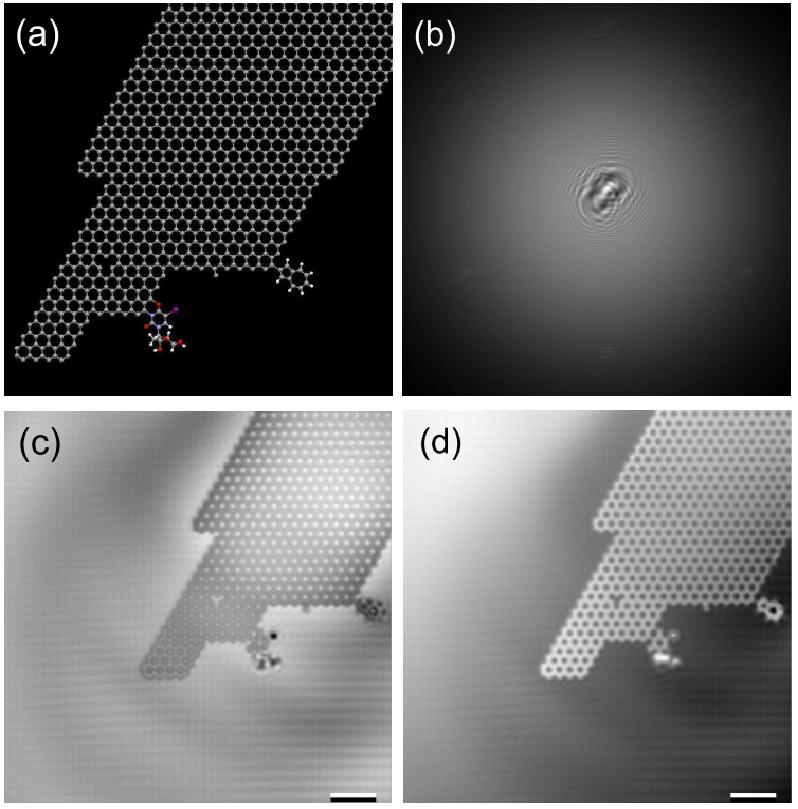}
}\caption{(a) Molecular model of a graphene piece featuring a vacancy defect (marked by an arrow) and adsorbed molecules (azulene and idoxuridine, respectively). (b) Simulated LEEPS hologram of the sample at 200 eV electron energy.  Reconstructed amplitude (c) and phase (d) image of the sample.  The scale bar in (c) and (d) is 1nm.} 
\label{fig3}
\end{figure}

In \fig{fig3} we show the molecular model (a), hologram (b) and reconstructions (c,d) of a graphene ribbon, in a region where it tapers down to a width of a few atoms. The hologram was simulated assuming an electron energy of 200 eV, source-object distance $d=$100 nm, source-screen distance $L=$1 cm, and a detector (experimentally, a micro channel plate detector) with 12 micron pore size. The honeycomb lattice is recovered well in both intensity and phase reconstructions. As in can be seen in the reconstructed images, lattice defects (e.g. vacancies, etc), and chemisorbed molecules on graphene edge can also be imaged by this method. Not shown here, but also amenable to imaging are substitutional impurities, physisorbed atoms and molecules, which are discussed below.   

In order to explore elemental contrast we further proceed to investigate to what extent atoms of different types adsorbed on graphene can be resolved and differentiated by hologram reconstruction. To this end we simulated the imaging of the graphene sample above on which individual atoms N, O, Ca, S, Zn, Au are adsorbed at different location. The locations for adsorption were chosen in the hollow of the aromatic rings and at a distance from the graphene plane corresponding to van der Waals radii.  

\begin{figure}[tb]
\centerline{
\includegraphics[width= 1.0\linewidth]{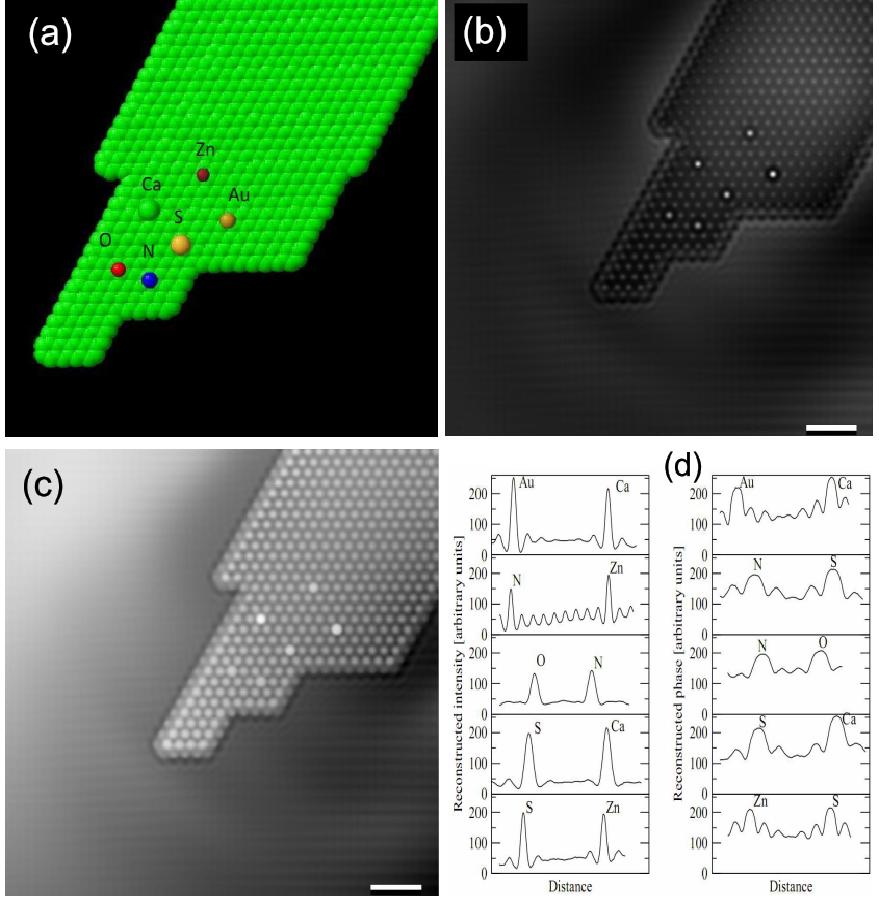}
}\caption{(a) Molecular model of a graphene piece featuring physisorbed atoms of different types (b) Reconstructed amplitude (b) and phase (c) images from the simulated hologram of the sample in (a) (scale bar is 1nm). (d) Sectional plots in the amplitude and phase cuts along lines through pairs of atoms indicated at each peak in the (b) and (c).	} 
\label{fig4}
\end{figure}

The molecular model and reconstruction images are shown in \fig{fig4}. The reconstructed amplitude values are in accordance with the elastic scattering cross sections of the atoms.  It is apparent that the scattering factor amplitudes for N, O are very similar, but substantially smaller than those of Ca, and Au, while Zn and S have values in between. Information on the atomic species can be extracted from both the magnitude and phase of the scattering factors, but also from the square of the scattering factor $|f(\theta )|^{2} $ (equal to the intensity of the reconstructed wave), which is a quantitatively related to the ESCS. To this end we plot in \fig{fig4} (d) linear plots of the intensity and phase corresponding to cuts along straight lines passing through pairs of atoms of different types. The peaks appearing in these plots are a measure of the reconstructed ESCS and phase, respectively of each atom. These results show that, for the setup parameters chosen here, one can actually discern between different atomic species based on their reconstructed scattering amplitude and phases. However, meaningful identification relies on the predetermined knowledge of the scattering factors, i.e. on correct theoretical predictions. One must also account for the subtleties below.

First, we must point out that an exact interpretation of these results must account for the fact that the scattering factors are highly anisotropic, see \fig{fig2}(b). As a consequence, the reconstructed scattering factors are in fact average values over the solid angular range spanned by the detector. It is therefore not possible to recover the exact scattering factors by reconstructing the whole hologram. 

It is however conceivable to extract more information about the angular form of the scattering factors by noting that the hologram can be sectioned into many circular segments corresponding to as many scattering angle intervals. Here we assume that the source-sample line is perpendicular to and aligned with the center of the detector. Given the fact that the size of our field of view is so much smaller than that of the detector, the scattering angle corresponding to any pixel of the detector is almost identical for all the atoms in the sample. Thus, by isolating and reconstructing such individual partial ``ring'' hologram, we are able to effectively measure the scattering factors at given scattering angles. In other words, by cutting out and reconstructing a series of ring holograms with different opening angles, the partial holograms act as effective bins for building histogram of the scattering factor amplitude and phase as functions of the scattering angle. However, there are limitations to the method: in order for such a partial hologram to provide enough information for an accurate reconstruction, it has to contain a minimum number of Fresnel fringes, which limits the actual width of each such ring and therefore the number of angular bins yielding the angular resolution of the measurement. Nonetheless, the method might be worth pursuing in detail as it provides - to our knowledge - the only way of actually measuring the \textit{complex phase} of the scattering factors at low electron energies.

\begin{figure}[tbh]
\centerline{
\includegraphics[width= 1.0\linewidth]{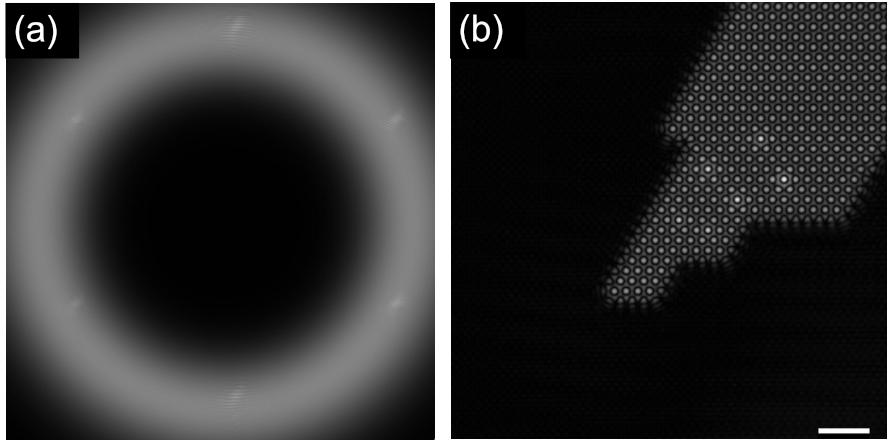}
}\caption{(a) Partial ring-shaped hologram obtained by multiplying the whole hologram by a radial Gaussian envelope centered at a scattering angle of 30 degrees and of a decay width of 7 degrees. (b) Amplitude reconstruction of the partial hologram (scale bar is 1 nm).} 
\label{fig5}
\end{figure}

In \fig{fig5} we show reconstructions of the same graphene sample obtained from a partial hologram obtained by applying a Gaussian envelope of the form $\exp [-(\gamma -\gamma _{0} )^{2} /\Delta \gamma ^{2} ]$ to the simulated whole hologram in \fig{fig3}(b). Here, $\gamma$ is the angle between the optical axis and line joining the source with a given pixel on the screen. The resulting partial hologram is shown in \fig{fig5}(a) and the amplitude reconstruction is shown in \fig{fig5} (b). The level of noise is greatly reduced by this procedure, but we note that the compositional contrast is also reduced. As the ring diameter is increased -- corresponding to reconstructions at increasing scattering angles - the contrast between the light and heavy atoms disappears. This fact can be attributed to the fact that the scattering factors for these atoms become much closer to each other at higher angles (see \fig{fig2}(b)).  We noted that phase images from partial holograms suffer from considerable more noise as compared to the full hologram reconstruction.

We also simulated holograms and their reconstructions for heterogeneous sample, containing a variety of atomic species. Some practically feasible samples are stiff polymer chains spanning across a sample holder (e.g. a hole in holey carbon) in its  different conformations. Such polymers can be substituted and functionalized with different chemical groups and elements. Our simulations show that the contrast between light elements (C, N, O) is negligible. The hydrogen atoms appear very ghostly, due to their scattering factors being much smaller than that of other species. However, we found that the substitution of slightly heavier elements, such as Si, P, S, Ca, offers substantial contrast with the rest of the polymer chain. Thus, such elements can be envisioned as stains for our microscopic method, and they are specific to the low electron energy range in which we work.

\section{Conclusion and Outlook}

Elemental contrast is amenable by LEEPS holography due to pronounced variations of atomic scattering factors with the atomic number at low electron energy (below 300 eV). We carried out atomistic simulation of in-line transmission hologram formation by elastic electron scattering at low electron energies, and we digitally reconstructed the resulting holograms. For a source close to an ideal one (single atom emission, large coherence width), the procedure shows the possibility of obtaining discernible elemental contrast with atomic resolution. The sensitivity of the elemental contrast is better in phase images than in amplitude images. This is due to the fact that on average the scattering phase difference between light and heavy elements is greater at high spatial frequencies, where the high resolution details are found. The scattering amplitude differences between light and heavy elements decrease with increasing spatial frequency. 

At this stage, absolute values for the amplitude and phase imaging are not feasible mainly due to the finite beam opening angle and/or the finite size of the detector and also due to a practical restriction of electron energies to a range from 50 to 200 eV producing wavelengths greater than 0.1 nm. Aiming for smaller wavelengths (higher energies) would likely increase the emission area on our source and also the incoherent source size, which would again reduce our resolution. However, the method produces good relative contrast between atoms of different types in both amplitude and phase images, consistent with the individual atomic scattering factors. The rule of thumb for achieving the maximum resolution is then: optimize the trade-off between the source size (increases with electron energy) and the wavelength (decreases with electron energy). From our estimates, it appears that an optimum is achieved at an electron energy around 100 eV.

For atoms adsorbed on graphene, we obtain contrast by reconstructing objects in planes parallel to the graphitic plane. Thus, any ultrathin, free standing structure becomes amenable to amplitude and phase imaging by this method, including static electric and magnetic fields associated with nanostructures. An advantage of our method over doing microscopy of magnetic fields at high electron energies (such as off-axis holography in a transmission electron microscope) is the greater sensitivity for phase shifting at low energy due to a greater interaction constant, which scales as the wavelength. Graphene, a nanostructure of current fundamental and technological interest, appears naturally as an ideal sample for this form of holographic microscopy. It also appears to be an ideal substrate for imaging atoms, molecules, and nanostructures by this technique. We can use the current method to study the structure of graphene edges, vacancy, impurity, and topological defects, adsorbed entities at the edge or on the surface of graphene membranes. 

Although here we present results only for the transmission mode, the reflective mode is also possible as outlined  in \cite{spence1997low,spence1999introduction}. Numerical simulations of electron holograms of various samples presented here show the possibility of attaining analytical microscopy with atomic resolution by lensless low energy electron holography. Further theoretical and experimental work is under way in order to explore the potential and limits of the LEEPS holographic microscopy method.

\ack{This work is supported by iCORE, the Alberta Ingenuity Fund and the National Research Council of Canada.}

\section*{References}



\newpage


\end{document}